\documentclass[12pt,preprint]{aastex}
\usepackage{}
\def\beq{\begin{equation}}
\def\enq{\end{equation}}
\def\ms{$M_\odot$}
\newcommand{\pfrac}[2]{\left(\frac{#1}{#2}\right)}

\shorttitle{}
\shortauthors{}
\begin{document}

\title{Hard X-ray emissions from Cassiopeia A observed by INTEGRAL}

\author{Wei Wang$^1$, Zhuo Li$^{2}$}
\affil{$^1$ National Astronomical Observatories, Chinese Academy of Sciences,
                 20A Datun Road, Chaoyang District, Beijing 100012, China;
                 wangwei@bao.ac.cn \\
      $^2$ Department of Astronomy and Kavli Institute for Astronomy and Astrophysics, Peking University, Beijing 100871, China; zhuo.li@pku.edu.cn
       }

\begin{abstract}
Cassiopeia A (Cas A) as the nearby young remnant of a core-collapse
supernova is the best candidate for astrophysical studies in
supernova explosion and its environment. We studied hard X-ray
emissions from Cas A using the ten-year data of INTEGRAL observations,
and first detected non-thermal continuum emission from the source up to 220 keV. The $^{44}$Ti line emissions at 68 and 78 keV are confirmed by our observations with a
mean flux of $\sim (2.2\pm 0.4)\times 10^{-5}$ ph cm$^{-2}$
s$^{-1}$, corresponding to a $^{44}$Ti yield in Cas A of $(1.3\pm
0.4)\times 10^{-4}$ \ms. The continuum emission from 3 --
500 keV can be fitted with a thermal bremsstrahlung of $kT\sim
0.79\pm 0.08$ keV plus a power-law model of $\Gamma \sim 3.13\pm
0.03$. The non-thermal emission from Cas A is well
fitted with a power-law model without a cutoff up to 220 keV. This
radiation characteristic is inconsistent with the
diffusive shock acceleration models with the remnant shock velocity of only 5000km s$^{-1}$. The central compact object in Cas
A cannot contribute to the emission above 80 keV
significantly. Some possible physical origins of the non-thermal emission
above 80 keV from the remnant shock are discussed.
We deduce that the asymmetrical supernova explosion scenario of Cas A is a promising scenario to produce high energy
synchrotron radiation photons, where a part of ejecta with the
velocity of $\sim 0.1c$ and opening angle of $\sim10^\circ$ can
account for the 100-keV emission, consistent with the "jet" observed
in Cas A.

\end{abstract}


\section{Introduction}

Supernova remnants (SNRs) are considered as the main sites to produce the high energy cosmic rays in the energies at least up to the so-called knee region of $\sim 3\times 10^{15}$ keV. The accelerated electrons and protons in the shock fronts can produce non-thermal emissions observed from radio to gamma-ray bands. The radio to X-ray emission comes from synchrotron radiation from accelerated electrons. The gamma-ray emission (GeV -- TeV) is either
caused by inverse Compton scattering by the same electrons that
cause synchrotron emission, or by pion production caused
by collisions of accelerated protons/ions with the background plasma. The more direct evidence for the efficient cosmic-ray acceleration comes
from detection of TeV gamma-rays in Galactic SNRs by ground-based Cherenkov Telescopes, e.g., HEGRA (Aharonian et al. 2001), MAGIC (Albert et al. 2007), HESS (Aharonian et al. 2004) and VERITAS (Humensky 2008). The non-thermal hard X-ray emissions (above $\sim 10$ keV) will be the other good way to probe the accelerating ability of SNRs. The first evidence for the shock acceleration of high-energy electrons up to 100 TeV comes from detecting the X-ray synchrotron emission in SN 1006 by ASCA (Koyama et al. 1995). Later OSSE onboard CGRO (The et al. 1996) and RXTE (Allen et al. 1997) observations also discovered hard X-ray tails up to $\sim 100$ keV in the supernova remnant Cas A, which suggested that this remnant can accelerate the electrons to energies at least $\sim 40$ TeV.

Cas A is the famous young supernova remnant ($\sim$ 330 years old) which is bright in all electromagnetic spectrum,
making it a unique laboratory for studying high-energy phenomena in SNRs. The distance of Cas A is determined at $\sim 3.4\pm 0.3$ kpc (Reed et al. 1995). Cas A is the first SNR detected in TeV bands, first by HEGRA (Aharonian et al. 2001), and confirmed by MAGIC (Albert et al. 2007) and VERITAS (Humensky 2008) with a photon index of $\sim 2.4\pm 0.3$. Recently FERMI/LAT also detected the GeV emission from Cas A, with a photon index of $\Gamma\sim 2.0\pm 0.1$ between 0.5 and 50 GeV (Abdo et al. 2010). Non-thermal hard X-ray emission from Cas A was also reported by OSSE (from 40 -- 120 keV, The et al. 1996) and RXTE (2 -- 60 keV, Allen et al. 1997) observations. Above 10 keV, the spectrum of Cas A can be fitted a power-law model of the photon index $\Gamma\sim 3.04$ (Allen et al. 1997). Suzaku observations of Cas A also detected the hard X-ray emission up to 40 keV (Maeda et al. 2009), whose spectrum from 3 -- 40 keV can fitted with a thermal bremsstrahlung and a power-law model with a photon index of $\Gamma\sim 3.06\pm 0.05$.

Hard X-ray observations on Cas A can also study the hard X-ray lines at 67.9 and 78.4 keV coming from the decays of radioactive $^{44}$Ti. $^{44}$Ti is a short-lived radioactive isotopes with a half life of 59
years (Admad et al. 2006). In theories, the most plausible cosmic environment for
production of $^{44}$Ti is the $\alpha$-rich freeze-out from
high-temperature burning near the nuclear statistical equilibrium
(Woosley et al. 1973; Timmes et al. 1996). So $^{44}$Ti should be produced in supernova explosions. $^{44}$Ti gamma-rays reflect the current rate of supernovae (SNe) due to
its short decay time scale, and probe the
inner regions of core-collapse SNe. Its yield also depends on the type of SNe. Solar metallicity Type
II and Type Ib supernova standard models indicate $\sim (3-6)\times
10^{-5}M_\odot$ yields of $^{44}$Ti (e.g. Woosley \& Weaver 1995;
Rauscher et al. 2002; Limongi \& Chieffi 2003).

There are three $\gamma$-ray lines which can be used to detect the decay of
$^{44}$Ti: the 67.9 and 78.4 keV lines from the $^{44}$Sc
de-excitation cascade and the 1157 keV line as $^{44}$Ca decays to
its stable ground state. The three lines are emitted with efficiencies of $87.7\%$ at 67.9 keV, $94.7\%$ at 78.4 keV, and $99.9\%$ at 1157 keV. Analysis of COMPTEL data supported a 5$\sigma$
detection of $^{44}$Ti in Cas A at $(4.2\pm 0.9)\times 10^{-5}\ \mathrm{photon\
cm^{-2}\ s^{-1}}$ in the 1157 keV line (Iyudin et al. 1994). The OSSE observations also marginally detected the $^{44}$Ti signal in Cas A around 60 -- 90 keV (The et al. 1996), giving a mean flux of $\sim (1.9\pm 1.6)\times 10^{-5}\ \mathrm{photon\ cm^{-2}\ s^{-1}}$ for the two low energy lines.  The BeppoSAX Phoswich Detector System (PDS) detection of the two blended hard X-ray lines
at 67.9 and 78.4 keV gave a $^{44}$Ti flux of $\sim (1.9\pm 0.8)\times 10^{-5}\ \mathrm{photon\
cm^{-2}\ s^{-1}}$ (Vink et al. 2001). Early INTEGRAL/IBIS observations detected two hard X-ray lines, and determined the $^{44}$Ti flux of $\sim (2.2\pm 0.5)\times 10^{-5}\ \mathrm{photon\ cm^{-2}\ s^{-1}}$ only using IBIS data, and an average flux of $\sim (2.5\pm 0.3)\times 10^{-5}\ \mathrm{photon\ cm^{-2}\ s^{-1}}$ by combining COMPTEL, BeppoSAX PDS and IBIS measurements (Renaud et al. 2006). Recent imaging observations of $^{44}$Ti in Cas A by NuStar suggested the asymmetry in this core-collapse supernova (Grefenstette et al. 2014), and the reported line flux at 68 keV is about $\sim (1.51\pm 0.31)\times 10^{-5}\ \mathrm{photon\ cm^{-2}\ s^{-1}}$.  The detected $^{44}$Ti flux deduced a yield range of $(1-2)\times 10^{-4}$ \ms in Cas A, which is higher than the $^{44}$Ti yields predicted by simulations of the standard spherical explosion models. This leaves the $^{44}$Ti observations from Cas A as a puzzle, requesting further studies.

In this work, we collect all the available INTEGRAL observations around Cas A at present to carry out analysis. More observational data compared with previous work will help to search for the non-thermal hard X-ray emission above 100 keV. With a broad energy band spectrum of Cas A, we can constrain the hard X-ray spectral properties for the non-thermal continuum. Then the hard X-ray line fluxes of $^{44}$Ti will be measured with a better continuum fit. In the following section, the INTEGRAL observations and data analysis will be briefly described. The hard X-ray spectral properties and $^{44}$Ti line features will be studied in \S 3. Finally, in \S 4 discussion on our results will be presented, specially, the implications of non-thermal emissions above 100 keV in Cas A will be discussed. We draw a brief conclusion in \S 5.

\section{INTEGRAL Observations and data analysis}

Main scientific objects of the INTEGRAL satellite include the nucleosynthesis via detection and fine spectroscopy of nearby supernova remnants (Winkler et al. 2003). Thus since its launch in 2002, INTEGRAL carried out the frequent pointing observations around the Cassiopeia region. In this work, we use the available archival data for the INTEGRAL observations
where Cas A is within $\sim 12$ degrees of the pointing
direction. The INTEGRAL observations showed a long-term frequent monitoring on Cas A from 2003 -- 2012. The total pointing observations have about 3000 science windows (duration of each science window is about 2000 s). The archival data used in our work are available from the INTEGRAL Science Data Center (ISDC). The analysis is done with the standard INTEGRAL off-line
scientific analysis (OSA, Goldwurm et al. 2003) software, ver. 10.

We mainly use the observational data obtained by the INTEGRAL Soft Gamma-Ray Imager (IBIS-ISGRI, Lebrun et al. 2003) which views the sky through a coded aperture mask. IBIS/ISGRI has a 12' (FWHM)
angular resolution and arcmin source location accuracy in the energy band of 15 -- 500 keV. The total on-source time of IBIS/ISGRI obtained in our analysis is about 6.8 Ms after excluding the bad data due to solar flares and the INTEGRAL orbital phase near the radiation belt of the Earth. JEM-X aboard INTEGRAL is the small X-ray telescope (Lund et al. 2003) which can be used to constrain the continuum spectrum below 30 keV combined with IBIS. JEM-X has a small field of view, which can only detect the source with the off-source angle below $\sim 5^\circ$. The total observational time obtained by JEM-X is only about $\sim 0.7$ Ms.

Individual pointings in collected IBIS and JEM-X data processed with OSA 10 were mosaicked to create the sky images for the source detection in the given energy ranges. In Fig. 1, the sky images in four energy bands obtained by JEM-X and IBIS are displayed. JEM-X detected Cas A in the range of 3 -- 10 keV with a significance level of $\sim 58.2\sigma$. For the higher energy ranges, IBIS detected Cas A with the significance levels of $\sim 54.8\sigma$, $14.9\sigma$ and $7.3\sigma$ in three energy bands: 20 -- 60 keV, 60 -- 90 keV and 90 -- 200 keV, respectively.

In addition, we also studied hard X-ray flux variation of Cas A from 2003 -- 2011 detected by the INTEGRAL missions. In Fig. 2, The normalized hard X-ray flux of Cas A versus observed time is presented in three energy bands: 3 -- 10 keV; 18 -- 35 keV and 35 -- 60 keV. The hard X-ray fluxes show no significant changes (within $\sim 20\%$) over nearly ten years. Though large error bars exist, there still may exist differences in flux variations among three energy ranges. In low energy band of 3 -- 10 keV, the flux in 2009 -- 2012 reduced to about 80$\%$ of that in 2003; while in high energy bands of 18 -- 35 keV and 35 -- 60 keV, the average changes of fluxes in two time intervals are generally blow $10\%$. So that the higher hard X-ray emissions in Cas A might decrease slowly relative to the lower X-ray bands.

\begin{figure*}
\centering
\includegraphics[angle=0,width=15cm]{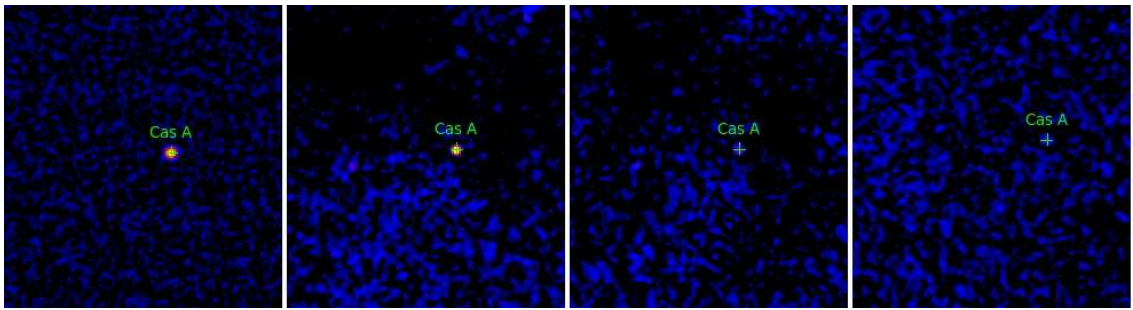}
\caption{Significance mosaic maps around the supernova
remnant Cas A in Equatorial J2000 coordinates as seen (panels from left to right) with INTEGRAL/JEMX in the range of 3 -- 10 keV and
INTEGRAL/IBIS in three energy bands: 20 -- 60 keV, 60 -- 90 keV and 90 -- 200 keV. Cas A was detected by JEM-X with a significance level of 58.2$\sigma$ and by IBIS with the significance levels of $\sim 54.8\sigma$, $14.9\sigma$ and $7.3\sigma$ in three energy bands, respectively. }
\end{figure*}

%\begin{figure}
%\centering
%\includegraphics[angle=0,width=9cm]{casa_count.eps}
%\includegraphics[angle=0,width=9cm]{casa_cts2b.eps}
%\caption
%{The hard X-ray count rates in the bands of 20 -- 60 keV detected by IBIS from 2003 -- 2011 for Cas A. No long-term variations of hard X-ray %flux are detected by INTEGRAL. }
%\end{figure}

\begin{figure}
\centering
\includegraphics[angle=0,width=14cm]{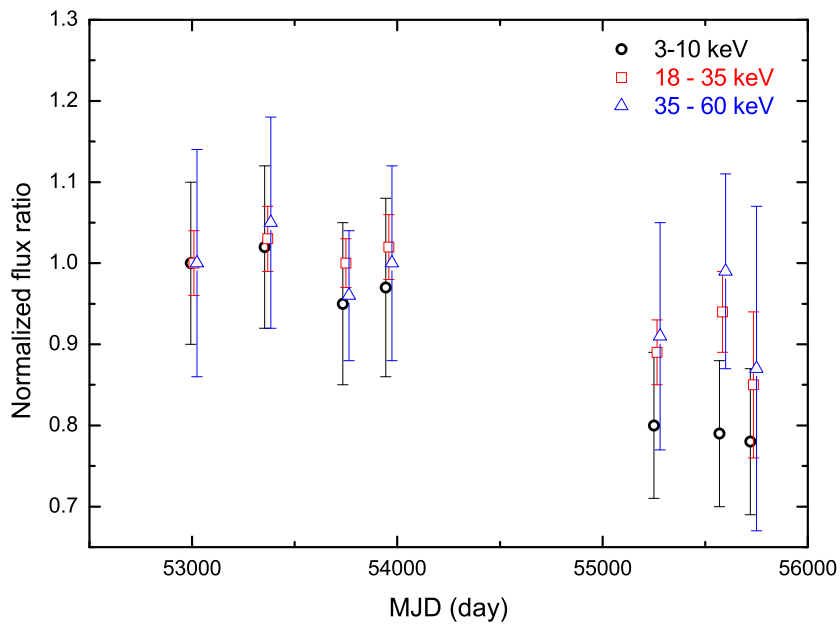}
\caption
{The detected flux variations from 2003 -- 2012 for Cas A in three X-ray bands: 3 -- 10 keV; 18 -- 35 keV; 35 -- 60 keV. The flux of Cas A observed in 2003 is normalized to 1. For clarity, we have shifted the data points of two bands (18 -- 35 keV and 35 -- 60 keV) to the right, separately. }
\end{figure}

\section{Hard X-ray spectral properties of Cas A}

In this section, we analyzed the overall spectrum from 3 -- 500 keV of Cas A with all available data. In Fig. 3, the extracted hard X-ray spectra from 3 -- 35 keV obtained by JEM-X and 18 -- 500 keV by IBIS are shown. Generally, the X-ray radiation from young supernova remnant Cas A should include both thermal emission component and non-thermal emission component. We have tried different continuum spectral models to fit the data, then checked which model can best fit the INTEGRAL data. In addition, if this count rate spectrum is only fitted by a simple power-law model (with $\Gamma\sim 3.1$), three line features can be clearly found in the residuals: strong Fe K$\alpha$ line around 6.5 keV; two hard X-ray lines at $\sim 68$ and 78 keV attributed to the decay of $^{44}$Ti. Thus, we showed the fitted reduced $\chi^2$ values (Table 1) by using four continuum models plus the iron line at 6.5 keV: a simple power-law model; a thermal bremsstrahlung plus a power-law model; a cool thermal bremsstrahlung plus a hot bremsstrahlung; a thermal bremsstrahlung plus an exponential decay. Comparing the different fitted reduced $\chi^2$ results in Table 1, the thermal bremsstrahlung plus a power-law model gave a best fit to the INTEGRAL data, and the other continuum models can be rejected with much larger reduced $\chi^2$.

\begin{table}
%\tabletypesize{\scriptsize}

\caption{Derived reduced $\chi^2$ values of fitting the INTEGRAL data points of Cas A by using different continuum models.  }
% \setlength{\tabcolsep}{1.0mm}
%\tablewidth{0pt}
\begin{center}
\begin{tabular}{l l}
\hline \hline
Model &  Reduced $\chi^2/d.o.f$    \\
\hline
simple power-law model $+$ iron line & 5.3811/69 \\
thermal bremsstrahlung plus a power-law model $+$ iron line & 2.1960/67 \\
a cool thermal bremsstrahlung plus a hot bremsstrahlung $+$ iron line &  5.7856/67 \\
thermal bremsstrahlung plus an exponential $+$ iron line & 6.1651/67 \\
\hline
\end{tabular}
\end{center}
\end{table}

Then the hard X-ray continuum was fitted with a thermal bremsstrahlung plus a power-law model, giving $kT\sim 0.79\pm 0.08$ keV and $\Gamma \sim 3.16\pm 0.03$. The thermal emission component fraction is about $40\%$ at 3 keV, and $\sim 13\%$ at 4 keV. The iron line feature around 6.5 keV was fitted with a Gaussian line profile. We found a line centroid energy of $\sim 6.61\pm 0.04$ keV, with a line width of $\sim 0.25\pm 0.16$ keV, and a line flux of $\sim (4.7\pm 0.2)\times 10^{-3}$ ph cm$^{-2}$ s$^{-1}$. The derived line parameters of the Fe K$\alpha$ feature are well consistent with those obtained by Suzaku observations (Maeda et al. 2009). The power-law index of $\Gamma \sim 3.1$ above 10 keV derived from the INTEGRAL data is consistent with the previous measurements by OSSE (The et al. 1996), RXTE (Allen et al. 1997), BeppoSAX (Favata et al. 1997), Suzaku (Maeda et al. 2009) and early INTEGRAL/IBIS data (Renaud et al. 2006).

In the top panel of Fig. 3, the spectrum was fitted with the continuum model with a thermal bremsstrahlung plus a power-law model, and an Fe K$\alpha$ line at $\sim 6.6$ keV. Two emission line features at $\sim 68$ and 78 keV appear obviously in the residuals. In the bottom panel of Fig. 3, we fitted the spectrum with a thermal bremsstrahlung plus a power-law model, and three gaussian line profiles: the Fe K$\alpha$ line, and two $^{44}$Ti decaying lines. For the two hard X-ray lines of $^{44}$Ti, we have fixed their energies at $\sim 67.9$ and 78.4 keV respectively in the fittings, and the line width is set to be zero due to the low spectral resolution of IBIS/ISGRI around 70 keV ($\sim 8\%$). In addition,  since the flux of the line at 67.9 keV is about $93\%$ of that of 78.4 keV, to improve the statistical significance of hard X-ray line detection, we also fix the line flux ratio during the fitting: $F_{68}=0.93F_{78}$. The best fitted spectral parameters by fitting the overall spectrum from 3 -- 500 keV of Cas A are presented in Table 2. The average line flux of $^{44}$Ti lines is determined at $F_{78}\sim (2.2\pm 0.4)\times 10^{-5}$ ph cm$^{-2}$ s$^{-1}$ according to the present measurements. The measured $^{44}$Ti line flux range is well consistent with the previous BeppoSAX (Vink et al. 2001) and IBIS (Renaud et al. 2006) measurements, and still lower than the COMPTEL results (Iyudin et al. 1994).

\begin{figure}
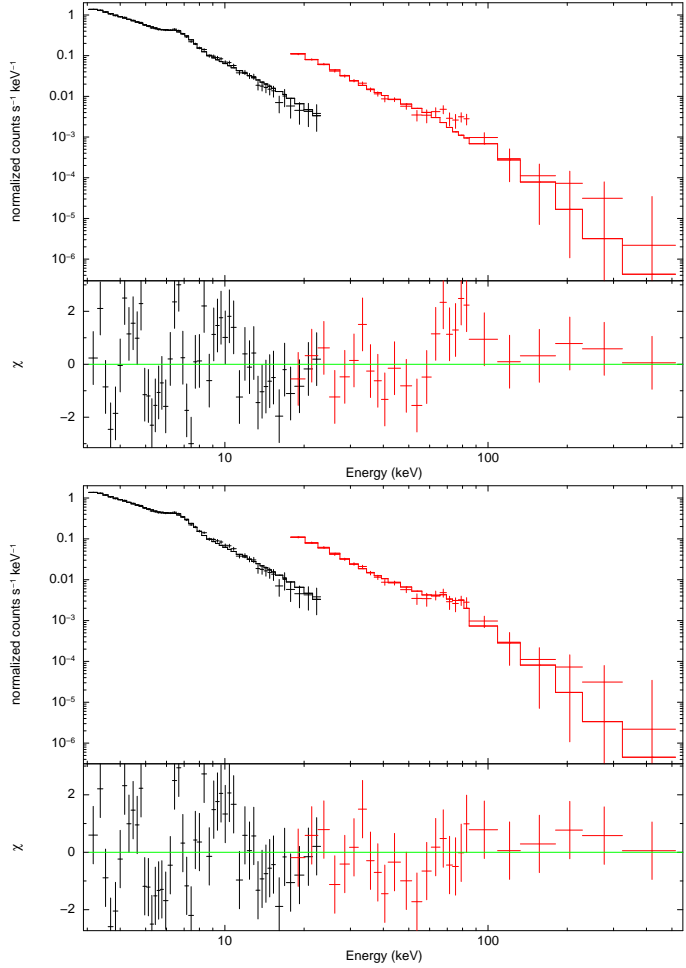

\centering
\includegraphics[angle=-90,width=9cm]{casa_brempow_fe.eps}
\includegraphics[angle=-90,width=9cm]{casa_44ti.eps}
\caption{{\bf Top:} The spectrum of Cas A in the hard X-ray band of 3 -- 500 keV. The spectrum is fitted with a thermal bremsstrahlung of $kT\sim 0.79\pm 0.08$ keV plus a power-law model of $\Gamma \sim 3.16\pm 0.03$, with an Fe K$\alpha$ line at $\sim 6.61\pm 0.04$ keV (reduced $\chi^2=2.1960$ (67 $d.o.f.$)). The continuum flux of 3 -- 200 keV is $\sim (6.3\pm 0.2)\times 10^{-10}$ erg cm$^{-2}$ s$^{-1}$. Two hard X-ray line features at $\sim 68$ keV and 78 keV are clearly seen in the residuals. {\bf Bottom:} Two $^{44}$Ti emission lines at $\sim 67.9$ and 78.4 keV are added to fit the overall spectrum of Cas A. See the text and Table 2 for details. }
\end{figure}

\begin{table}
%\tabletypesize{\scriptsize}

\caption{The hard X-ray spectral properties of Cas A by fitting the overall spectrum from 3 -- 500 keV with the continuum model of the thermal bremsstrahlung plus a power-law component.   }
% \setlength{\tabcolsep}{1.0mm}
%\tablewidth{0pt}
\begin{center}
\begin{tabular}{l c l}
\hline \hline
Continuum &  &    \\
\hline
Thermal bremsstrahlung & Power law & \\
$kT$ (keV)  &  $\Gamma$  & Flux (3 -- 200 keV, $10^{-10}$ erg cm$^{-2}$ s$^{-1}$) \\
0.81$\pm 0.08$ & 3.13$\pm 0.03$ & 6.3$\pm 0.2$ \\
\hline
Fe K$\alpha$ line &  &  \\
\hline
Energy (keV) & Width (keV) & Flux (10$^{-3}$ ph cm$^{-2}$ s$^{-1}$) \\
6.61$\pm 0.04$ & $0.25\pm 0.16$ & 4.7$\pm 0.2$ \\
\hline
$^{44}$Ti lines & & \\
\hline
Energy (keV) & Width (keV) & Flux (10$^{-5}$ ph cm$^{-2}$ s$^{-1}$) \\
67.9 (fixed) & 0(fixed) & 0.93$F_{78}$ \\
78.4 (fixed) & 0(fixed) & 2.2$\pm 0.4$ \\
\hline
Reduced $\chi^2$ ($d.o.f$) & 1.389 (66)  &  \\
\hline
\end{tabular}
\end{center}
\end{table}

Previous observations detected the non-thermal hard X-ray emissions in Cas A up to 100 keV by OSSE (The et al. 1996), RXTE (Allen et al. 1997), BeppoSAX (Favata et al. 1997), Suzaku (Maeda et al. 2009) and early INTEGRAL/IBIS data (Renaud et al. 2006). These previous detections showed that the hard X-ray emission of Cas A from 10 - 100 keV can be described by a power-law model of $\Gamma\sim 3.0-3.1$ or a hot bremsstrahlung model ($kT\sim 30$ keV). With accumulating nearly ten years of INTEGRAL data, we first detected the non-thermal X-ray emission extended to $\sim 220$ keV from Cas A (see Fig. 3). This non-thermal emission from 10 -- 220 keV is well consistent with a simple power-law of the photon index $\Gamma \sim 3.1$. No spectral cutoff is detected in hard X-ray bands. The hot bremsstrahlung model is rejected by the new INTEGRAL data (also see Table 1). The derived non-thermal continuum flux from 3 -- 200 keV is $\sim 5.7\times 10^{-10}$ erg cm$^{-2}$ s$^{-1}$ from Cas A (the power-law component), corresponding to a hard X-ray luminosity of $\sim 8\times 10^{35}$ erg s$^{-1}$ assuming a distance of 3.4 kpc.

\section{Discussion}

\subsection{$^{44}$Ti yield in Cas A}

With the measured hard X-ray line flux at 78 keV, the $^{44}$Ti yield in Cas A can be estimated as $M_{44Ti}\approx 4\pi d^2 44m_p \tau \exp(t/\tau) F_{78}$, where $d=3.4$ kpc is the distance of Cas A, $m_p$, the proton mass, $\tau=85$ yr, the characteristic time of the $^{44}$Ti decay chain, $t=335$ yr, the time since the explosion. The derived $^{44}$Ti yield in Cas A is $\sim (1.4\pm 0.3)\times 10^{-4}$ \ms.

\begin{figure}
\centering
\includegraphics[angle=0,width=13cm]{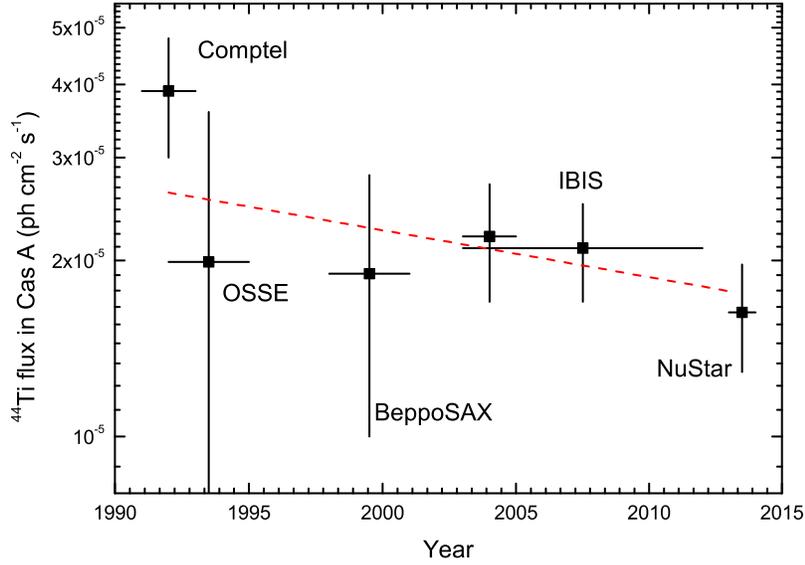}
\caption
{The detected $^{44}$Ti fluxes at 78 keV from 1992 -- 2014 for Cas A by different missions. We have cited the fluxes reported from the previous literatures: COMPTEL (Iyudin et al. 1994), OSSE (The et al. 1996), BeppoSAX (Vink et al. 2001), NuStar (Grefenstette et al. 2014) and the early IBIS result (Renaud et al. 2006). For the better comparison for different measurements in the plottings, we normalize the flux values to the flux levels at the 78 keV line by reducing the COMPTEL flux around the 1157 keV line by $5.2\%$ and increasing the NuStar flux at the 68 keV line by $7\%$ due to the differences in the branching ratios of the $^{44}$Ti lines. The red dashed line shows the $^{44}$Ti flux decaying curve if we assume the initial $^{44}$Ti yield of $1.3\times 10^{-4}$ \ms in Cas A, the explosion date around AD 1671 (Thorstensen et al. 2001), the distance of 3.4 kpc and the half life of $^{44}$Ti decay chain at 59 years.  }
\end{figure}

In the past twenty years, the measurements of three $^{44}$Ti lines in Cas A have been carried out by different missions. Here we collected all these reported $^{44}$Ti line fluxes together in Fig. 4. For the comparison in the fluxes by different
 measurements, we normalize the flux values to the flux levels at the 78 keV line, then reduce the COMPTEL flux around the 1157 keV line by $5.2\%$, and increase the NuStar flux at the 68 keV line by $7\%$. The observed $^{44}$Ti line flux decreased in the last twenty years, as expected for the decaying feature of the unstable $^{44}$Ti isotope. In Fig. 4, we also plotted the possible decaying curve of the $^{44}$Ti line flux in Cas A (red dashed line) by assuming that the initial $^{44}$Ti yield in Cas A is about $1.3\times 10^{-4}$ \ms, the explosion date is around AD 1671 (Thorstensen et al. 2001) with the distance of 3.4 kpc (Reed et al. 1995) and the half life of $^{44}$Ti decay chain at 59 years (Admad et al. 2006). The observed data points are generally consistent with the decaying curve, except that the COMPTEL data point is still above the curve. The 1157 keV from excited $^{44}$Ca$^*$ may have an additional nuclear de-excitation component originating from accelerated particles colliding with $^{44}$Ca$^*$ in the supernova remnant and swept-up interstellar medium (Siegert et al. 2015). However, the low energy lines at 68 and 78 keV come from the decay of the isotope $^{44}$Sc. $^{44}$Sc is short-lived, then no corresponding excitation of $^{44}$Sc will occur through such a cosmic ray process.

Generally people accepted that Cas A was formed by the explosion
of a massive star progenitor (possible type Ib), a 16 \ms single star (Chevalier \& Oishi 2003) to a Wolf-Rayet (WR) remnant of a very massive
($>60$ \ms ) precursor (Fesen \& Becker 1991). The standard spherical explosion models (e.g., Woosley \& Weaver 1995; Limongi \& Chieffi 2003) predict the $^{44}$Ti yield below $10^{-4}$ \ms, which contradicts with the observational constraints. Since in the standard
models of core-collapse supernova explosions, $^{44}$Ti and $^{56}$Ni are co-produced during the first stages
of the explosion, the high $^{44}$Ti may require that Cas A should be brighter due to a larger product of $^{56}$Ni. There exist some indications that its explosion energy was about $2\times 10^{51}$ ergs (Laming \& Hwang 2003), higher than the
canonical value of $10^{51}$ ergs. The other possibility is that the explosion of Cas A could be intrinsically asymmetric. And the sensitivity of the $^{44}$Ti production
to the explosion energy and asymmetries (Nagataki et
al. 1998) may explain the higher $^{44}$Ti/$^{56}$Ni ratio compared to standard spherical explosion models. This asymmetric explosion scenario is also supported by some observational evidence in the X-ray emitting ejecta of Cas A (Vink 2004; Hwang et al. 2004; Hwang \& Laming 2012). The recent direct imaging observations of the $^{44}$Ti emission in Cas A suggested an intermediate asymmetry in this core-collapse supernova (Grefenstette et al. 2014): the $^{44}$Ti is extended along the jet axis seen in X-rays (Hwang et al. 2004). Therefore Cas A may be a very special case of supernova explosions in the Galaxy, and produced by an asymmetric and/or a relatively more energetic explosion.

\subsection{Non-thermal power-law emission up to 220 keV in Cas A}

\begin{figure}
\centering
\includegraphics[angle=0,width=11cm]{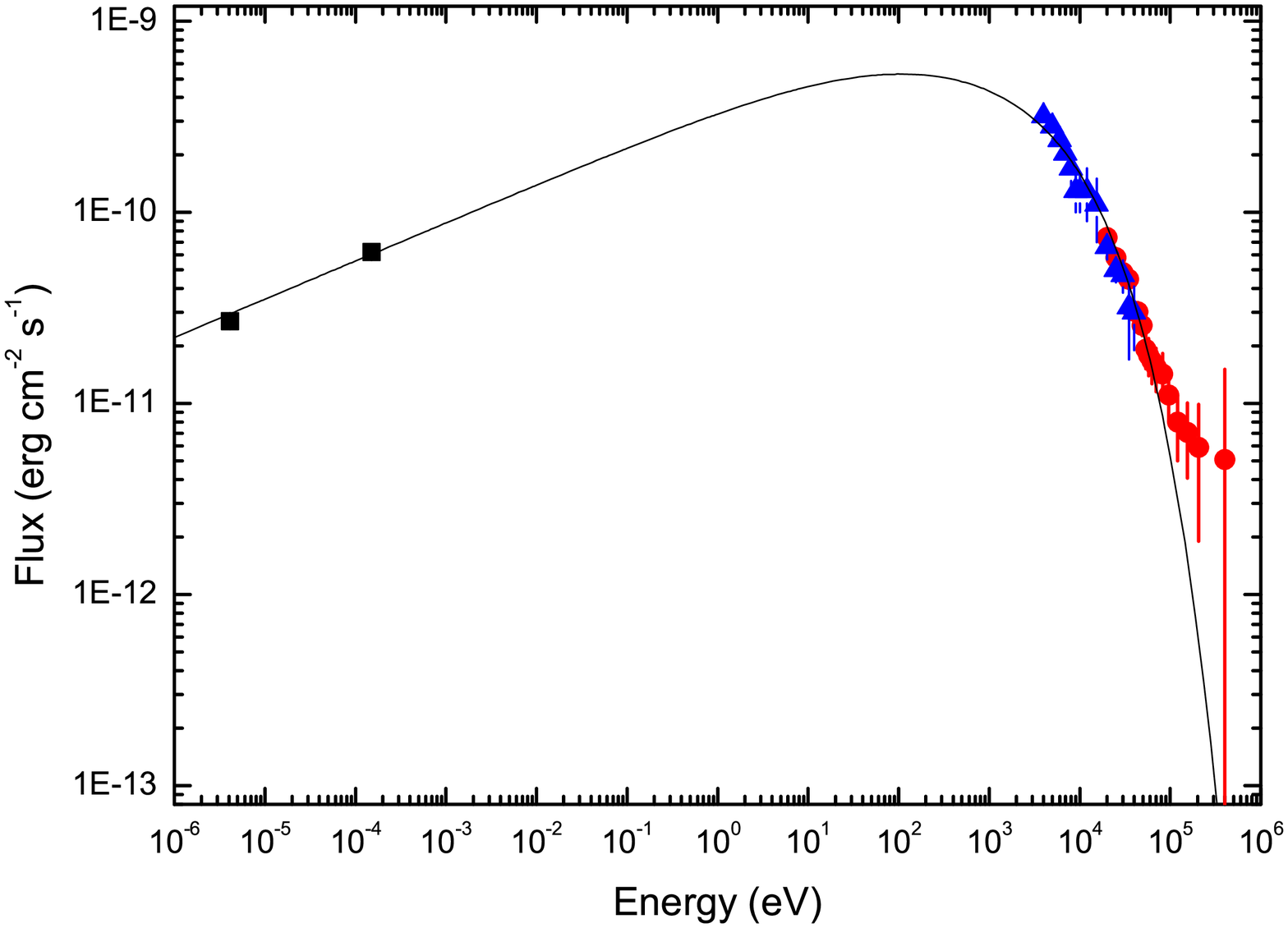}
\includegraphics[angle=0,width=11cm]{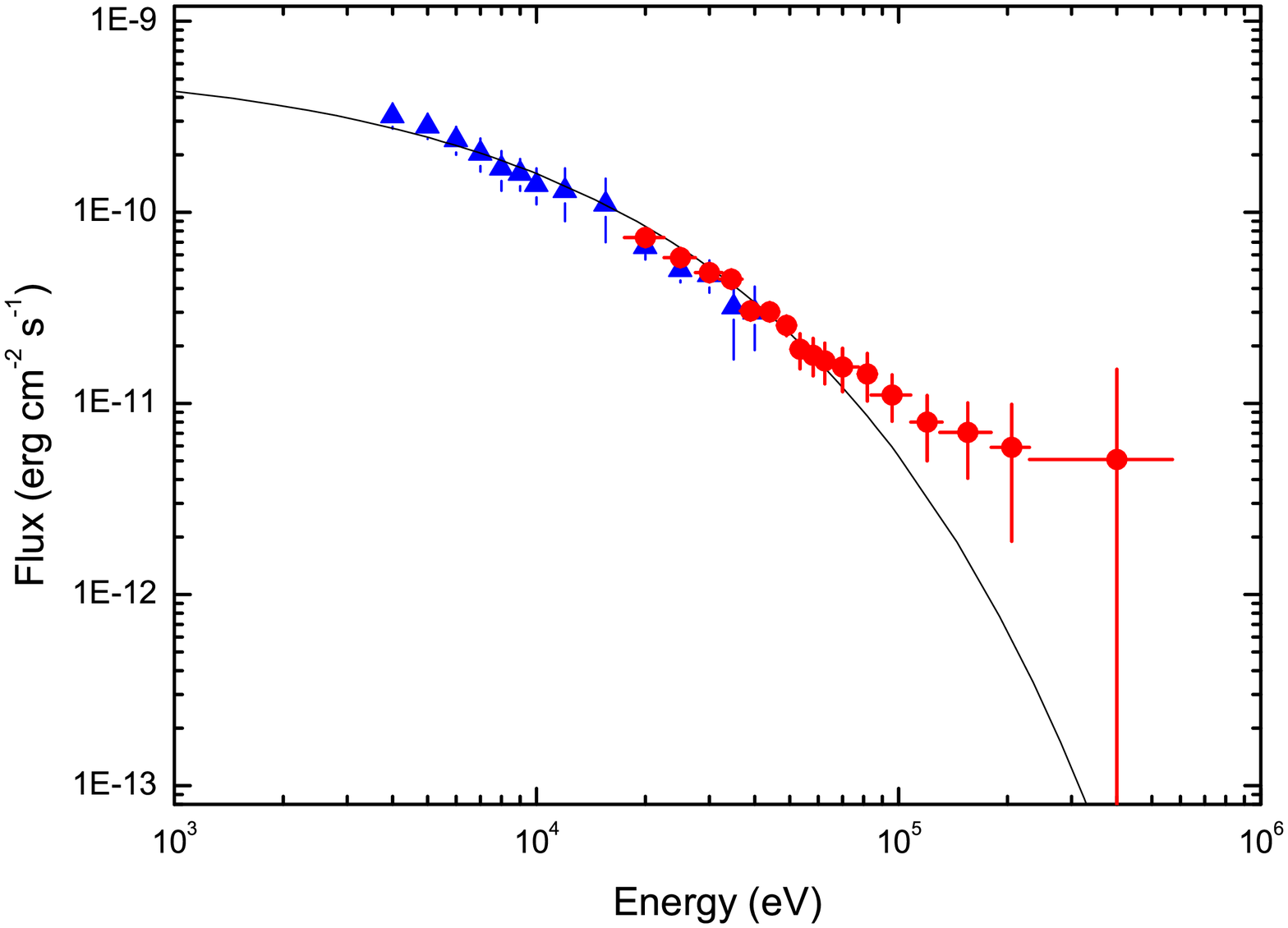}
\caption
{The observed spectra of Cas A from radio (black squares, from Artyukh et al. 1967) to X-ray bands including the Suzaku (blue triangles, from Maeda et al. 2009) and present IBIS (red circles) data points. The solid line is the synchrotron emission curve using the model of Zirakashvili \& Aharonian (2007). The top panel shows the data points and the fitting curve from the radio band to X-rays. The bottom panel shows the zoom-in part of the fittings in the X-ray bands from 1 -- 300 keV. The slow function derived by Zirakashvili \& Aharonian (2007) cannot fit the data points above 80 keV. }
\end{figure}

The non-thermal emissions from radio to X-ray bands in supernova
remnants are generally believed to be produced by the synchrotron
radiation of relativistic electrons accelerated in the remnant
shocks. Based on the diffusive shock acceleration (DSA) theory, the
produced synchrotron radiation spectrum shows a cutoff energy around
a few keV. The present observed hard X-ray emission by IBIS in Cas A
has a cutoff around a few keV, but the emission beyond the cutoff
energy shows a power-law extended up to 220 keV, which is
inconsistent with the DSA models. Recently Zirakashvili \&
Aharonian (2007; 2010) investigated the spectral shape of the
shock-accelerated electrons subject to synchrotron cooling in the
context of DSA theory, then they provided useful approximations for
the subsequent synchrotron spectral shape, which is a slowly
decreasing function other than sharp cutoff. We use this approximation (eq. 37 in
Zirakashvili \& Aharonian 2007) to fit the data points from the
radio to hard X-ray bands observed in Cas A (see Fig. 5). This slow
function can well fit the data points from radio to X-ray bands up
to 70 keV, however, the data points above 80 keV derived by IBIS
cannot be described by the approximation. Indeed the slower
decrease at higher energy band (hinted in Fig. 2) may suggest the hardening
spectral component at high energy.

\begin{figure}
\centering
\includegraphics[angle=0,width=11cm]{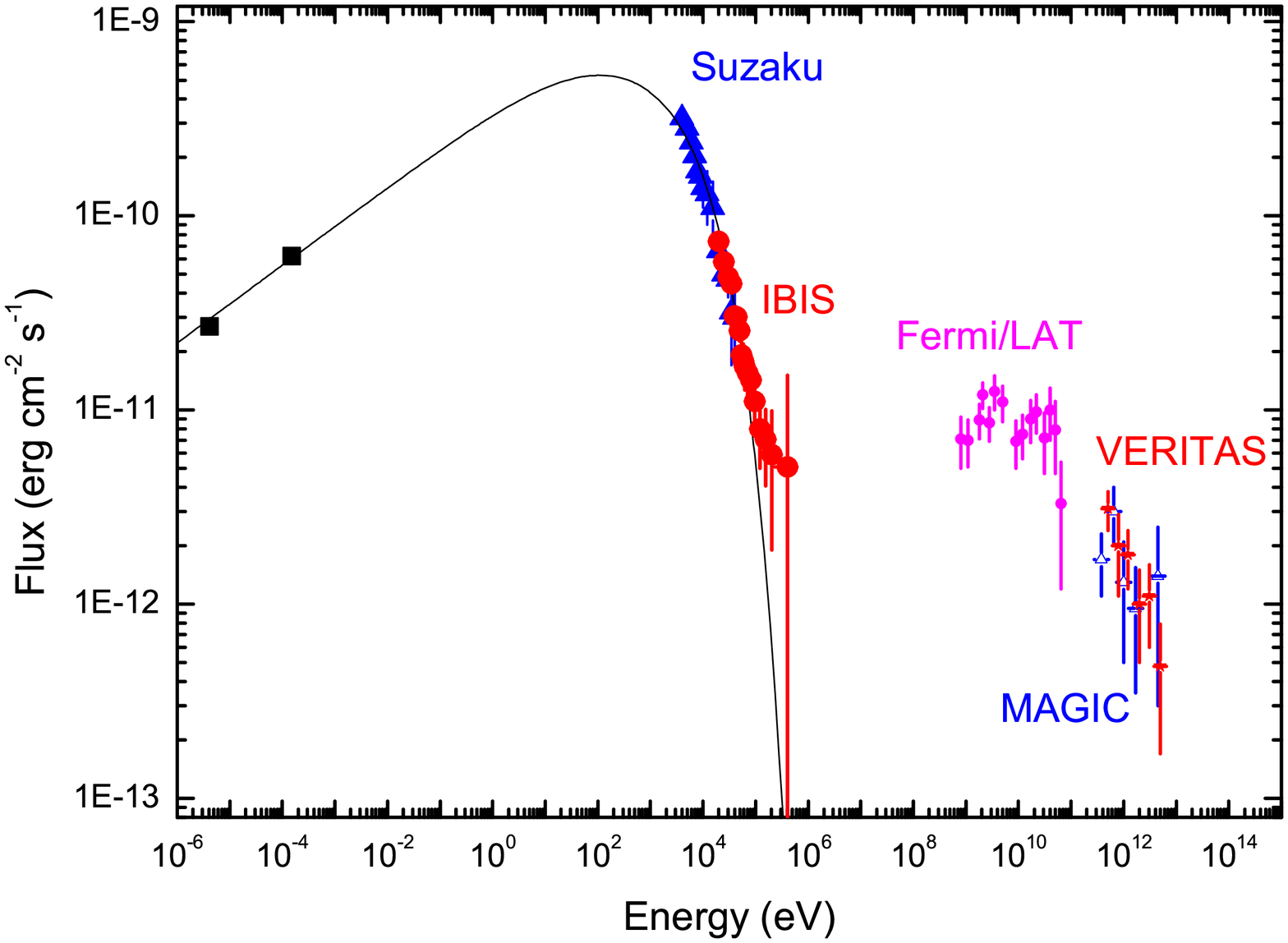}
\caption
{The observed multi-wavelength spectra of Cas A from radio, X-ray to GeV-TeV gamma-ray bands. IBIS data points is from the present work. The other points are taken from the published literatures: radio (Artyukh et al. 1967); Suzaku (Maeda et al. 2009); Fermi/LAT (Abdo et al. 2010); VERITAS (Humensky 2008); MAGUC (Albert et al. 2007). The solid line is the synchrotron emission curve using the model of Zirakashvili \& Aharonian (2007). }
\end{figure}

What is the physical origin of the non-thermal emission from 80 --
220 keV? Cas A is thought to be produced by a type II explosion, then after the explosion, a neutron star or black hole should be left in the center. In the center of the remnant, a compact source is discovered in the first light observations of Chandra X-ray Observatory (Tanabraum 1999). Further deep observations find the spectrum of the point source in the band of 0.5-- 5 keV can be well fitted by a black body of $kT_{bb}\sim 0.5$ keV (Pavlov et al. 2000; Chakrabarty et al. 2001), which is consistent with the standard neutron star cooling. This spectrum is also explained by the fallback accretion model into a neutron star or a black hole. If we assume the compact object could be a young normal pulsar, the power-law fit can obtain a photon index of $\Gamma\sim 3.1$ (Pavlov et al. 2000; Chakrabarty et al. 2001) which is much softer than other known young pulsars ($\Gamma\sim 1.5-2$). Could the possible hard X-ray emission from the compact object explain the observed hard X-ray flux?

Thermal radiation models from the neutron star cooling or fallback accretion cannot produce the non-thermal emission above 80 keV, so we only consider the possible origin of an active pulsar for the point source in Cas A. The soft X-ray flux of the point source in Cas A is very weak, with $F_x\sim 5\times 10^{-13}$ erg cm$^{-2}$ s$^{-1}$ in the range of 1-- 5 keV (Pavlov et al. 2000; Chakrabarty et al. 2001). If the source has a power-law emission with the photon index of $\Gamma\sim 3$, the hard X-ray flux attributed by the pulsar candidate is only at the level of $\sim 10^{-14}$ erg cm$^{-2}$ s$^{-1}$ above 80 keV, much lower than the total observed flux in Cas A. The pulsar component could not significantly contribute to the non-thermal emission above 80 keV, except that we assume the photon index of non-thermal emission from the pulsar candidate in Cas A changing from 3 below 5 keV to around 1 above 10 keV. However, the hard spectrum of the photon index around 1 is not consistent with the observed non-thermal emission above 80 keV. In addition, NuStar has not detected the compact source in the range of 3 -- 79 keV (Grefenstette et al. 2014), suggesting that the compact source cannot become brighter above 5 keV.

Thus, the compact source will not contribute to the non-thermal
emission from 80 -- 220 keV significantly. In the following part, we
assume that the emission is due to synchrotron radiation from the
remnant shock, and discuss the requirement for producing 100-keV
emission.

The acceleration of electrons in the remnant shock suffers from
radiative energy loss. The maximum synchrotron photon energy where
the electron acceleration and synchrotron cooling times are equal is
$\epsilon_{\max}\sim3.8\xi^{-1}(v/5000\rm \,km\,s^{-1})^2 keV$
(e.g., Katz \& Waxman 2006), where $\xi\ge1$ is the ratio of the
diffusion coefficient to the Bohm diffusion one, and $v$ is the
shock velocity. As the bulk velocity of the shock in Cas A is
$v\sim5000\rm km\,s^{-1}$ (e.g., Fesen et al. 2006; Grefenstette et
al. 2014), the cutoff energy is at a few keV for the Bohm limit
$\xi\sim1$. The Cas A spectrum is consistent with a turnover around
a few keV. However, by the DSA theory the spectrum beyond the cutoff
is an exponential decay (e.g., Zirakashvili \& Aharonian 2007),
inconsistent with our result at $\ga100$ keV. Since the acceleration
is not expected to be faster than Bohm limit ($\xi\ll1$), the
100-keV emission cannot be produced by the bulk remnant shock of Cas
A.

One may argue that the postshock magnetic field may not be uniform.
There may be stronger magnetic field distributed in smaller spatial
region. Thus, when the electrons are sampling the high magnetic
field region they emit higher energy synchrotron photons. Since the
synchrotron photon frequency goes with magnetic field as
$\epsilon\propto B$, the $\sim100$keV emission compared to a few keV
cutoff needs a magnetic field contrast of $\sim30$ times, which may
be too large to imagine. The other option is that the magnetic field
increases downstream with the distance away from the shock front, so
that electrons may produce higher energy photons when flowing
downstream. But so far there is no support from theory and
observation for magnetic field increasing downstream.

One may also expect that if the relativistic electrons that are
responsible to the 100-keV synchrotron emission is not produced by
shock acceleration, then it is possible to avoid the limit to the
synchrotron photon energy. A possible high energy electron origin is
the hadronic process that the shock-accelerated cosmic rays interact
with the background medium and produce pions which decay quickly
into secondary electrons (or positrons), as well as neutrinos and
photons, i.e., $\pi^\pm\rightarrow e+\nu_e+2\nu_\mu$, and
$\pi^0\rightarrow2\gamma$. Because of the connection between the
secondary electrons and photons, there should be also a connection
between the secondary photons and the electrons-emitted synchrotron
radiation. Because the produced $\pi^+$'s, $\pi^-$'s, and $\pi^0$'s
have similar numbers and energy, and the secondaries approximately
share the primary energy, we have $dn_\gamma=dn_e$, $E_\gamma=2E_e$,
and hence $E_\gamma^2dn_\gamma/dE_\gamma=2E_e^2dn_e/dE_e$. Since the
synchrotron photon energy is $\epsilon\propto E_e^2$ and the
electrons rapidly lose energy by synchrotron radiation, we have
$\epsilon^2dn_{\rm
syn}/d\epsilon=(1/2)E_e^2dn_e/dE_e=(1/4)E_\gamma^2dn_\gamma/dE_\gamma$.
For the 100-keV synchrotron radiation the corresponding secondary
photon energy is $E_\gamma=2E_e=0.95(\epsilon/100\rm
keV)^{1/2}(B/10^{-5}G)^{-1/2}$PeV. In the broad band spectrum of Cas
A (Fig. 6), the 10-TeV flux is already more than one order of
magnitude smaller than the 100-keV flux,
$E_\gamma^2dn_\gamma/dE_\gamma({\rm 10 TeV})\ll\epsilon^2dn_{\rm
syn}/d\epsilon({\rm 100 keV})$. Since the spectrum decays fast
beyond TeV, the PeV flux is even lower. So the radiation at 100 keV
is impossible to be of secondary electron origin.

\subsubsection{Asymmetrical supernova explosion?}
A possible model for the hard X-ray emission may be asymmetrical
explosion where the ejecta velocity is angularly dependent, and the
100-keV emission can be produced by the highest velocity ejecta. For
the emission cutoff to reach $\epsilon_{\max}\sim100$ keV, the
required ejecta velocity is about $v\sim0.1c$ with $\xi\sim1$. Let
us estimate the angular distribution of the ejecta velocity that can
produce the observational high energy X-ray emission. Assume the
solid angle of ejecta with velocity $v$ follows a power law of
$d\Omega/dv\propto v^{-\alpha}$. We derive the index $\alpha$ below.

Consider an area element in the shock surface with a solid angle of
$\Delta\Omega$, velocity $v$, and radius of $r\sim vt$ with $t$ the
supernova age. The number of swept-up medium electrons is $\Delta
N\propto r^3\Delta\Omega\propto v^3\Delta\Omega$. The electrons are
accelerated and follow a power law distribution,
$dn_e/d\gamma\propto\gamma^{-s}$ ($\gamma>\gamma_m$). The postshock magnetic field and
minimum electron Lorentz factor are $B\propto\sqrt{\rho v^2}$
($\rho$ is the medium density), and $\gamma_m\propto v^2$,
respectively, and the cutoff frequency is $\nu_{\max}\propto v^2$.
Denote $\nu_c$ the synchrotron frequency that is emitted by
electrons with cooling timescale equal to the supernova age. Because
$\nu_c<\nu_{\max}$, the synchrotron flux at the cutoff frequency is
(Sari et al. 1998)
\begin{eqnarray}
   \Delta F_{\nu, \max}=\Delta F_{\nu_m}\pfrac{\nu_c}{\nu_m}^{-(s-1)/2}\pfrac{\nu_{\max}}{\nu_c}^{-s/2}.\nonumber
\end{eqnarray}
Here the characteristic frequencies are
$\nu_m\propto\gamma_m^2B\propto v^5$ and $\nu_c\propto
B^{-3}t\propto v^{-3}$ (Sari et al. 1998), and the peak flux is
$\Delta F_{\nu_m}\propto\Delta NB\propto v^4\Delta\Omega$. Inserting
these scaling into the above equation, we have $
  dF_{\nu,\max}/d\Omega\propto v^{3s/2}.
$ Taking $\nu_{\max}\rightarrow\nu$, we have the spectrum
\begin{eqnarray}
  \frac{dF_\nu}{d\nu}\propto\frac{dF_\nu}{d\Omega}\frac{d\Omega}{dv}\frac{dv}{d\nu}\propto  v^{3s/2-\alpha-1}\propto\nu^{\frac34s-\frac12\alpha-\frac12}\nonumber
\end{eqnarray}
and $
   F_\nu\propto \nu^{\frac34s-\frac12\alpha+\frac12}.
$ If the observed spectrum shows a photon index $\Gamma$, then $
   \alpha=3s/2+2\Gamma-1.
$ In Cas A, $\Gamma\sim3.1$, and $s\sim2.2$ , then we derive
$\alpha\sim8.3$.

Assume axis-symmetrical explosion, $d\Omega\sim2\pi\sin\theta
d\theta\propto d\theta^2$ (if $\theta\ll1$), then $\theta\propto
v^{-\frac{\alpha-1}2}$, or $v\propto\theta^{-\frac2{\alpha-1}}$. For
$\alpha\sim8.3$, the velocity profile is
\begin{eqnarray}
%  \theta\simeq\theta_0\pfrac v{v_0}^{-3.65}\\
  v\simeq v_0\pfrac \theta{\theta_0} ^{-0.27}\nonumber
\end{eqnarray}
In Cas A, the bulk ejecta with lowest velocity have
$\theta_0\sim\pi$ and $v_0\sim5000$km\,s$^{-1}$, then the ejecta
with velocity $v\ga10^4\rm km\,s^{-1}$ contain an angle of
$\theta\la10^\circ$.

The shock radius should roughly follows the velocity profile, $r\sim
vt\propto\theta^{-0.27}$. It should be noted that there is a ``jet"
observed in Cas A, with opening angle $\theta_{jet}\sim12.5^\circ$,
and transverse expansion velocity $v_{\rm
tran}\sim(1-1.4)\times10^4$km\,s$^{-1}$ (Fesen et al.2006), that is
consistent with the requirement to explain the photon index of the
100-keV emission. Because of the projection effect the true velocity
should be larger than the measured transverse velocity, and may
reach $\sim0.1c$ to produce the 100-keV emission.

\section{Conclusion and summary}

Cas A is studied with the near ten-year data of INTEGRAL observations. We first detect the non-thermal hard X-ray continuum from Cas A up to 220 keV. The spectrum of Cas A from 3 -- 500 keV is derived by the INTEGRAL/JEM-X and IBIS telescopes. The broad-band continuum spectrum can be fitted with a thermal component with a thermal bremsstrahlung model of $kT\sim 0.79\pm 0.08$ keV plus a power-law model of $\Gamma \sim 3.16\pm 0.03$. The Fe $K\alpha$ line is detected with the centroid energy of $\sim 6.61\pm 0.04$ keV, a line width of $\sim 0.25\pm 0.16$ keV, and a line flux of $\sim (4.7\pm 0.3)\times 10^{-3}$ ph cm$^{-2}$ s$^{-1}$. The $^{44}$Ti emission lines at 68 and 78 keV are confirmed by our observations. The $^{44}$Ti lines are detected with a significance level of $\sim 5.5\sigma$. The mean $^{44}$Ti line flux at 78 keV is $\sim (2.2\pm 0.4)\times 10^{-5}$ ph cm$^{-2}$ s$^{-1}$ which is consistent with the previous measurements. The derived $^{44}$Ti yield from the explosion of Cas A is $\sim (1.4\pm 0.3)\times 10^{-4}$ \ms according to the present measurements.

Cas A is detected in hard X-ray bands up to $\sim$ 220 keV. In
addition, the spectrum from 20 -- 220 keV shows a single power-law
feature without cutoff. This spectral feature (specially non-thermal
emission above 80 keV) is inconsistent with the present DSA
models. The central compact object in Cas A could not contribute to
the non-thermal hard X-ray emission significantly. This non-thermal
component above 80 keV should come from the synchrotron radiation in
the shock of the remnant. We have discussed the possible physical
origin of the non-thermal emission above 100 keV from the shock
waves. There may exist a higher magnetic field in some small region
in the remnant or the magnetic field might increase downstream with
the distance away from the shock front. However, there is no
support from theory and observations for the magnetic field spatial
variation by a large contrast of tens of times in the supernova
remnants. The other possible contribution comes from synchrotron
radiation of the secondary relativistic electrons which originate in
hadronic process, but this component is about one order of magnitude
lower than the observed hard X-ray flux in Cas A. Finally we discuss
the radiation from the asymmetrical supernova explosion which is
also supported by the $^{44}$Ti observations in Cas A. If the
fastest ejecta have a velocity of $\sim 0.1c$ and contain an angle
of $\la10^\circ$, their synchrotron radiation can produce the
non-thermal emission around 100 keV and explains the observed
spectral index. This is consistent with the other observation that
shows a ``jet" in Cas A.

\begin{acknowledgements}
We are grateful to the referee for the suggestions. This work is based on observations of INTEGRAL, an ESA project with
instrument and science data center funded by ESA member states. This work is supported by the National Natural Science Foundation of China
(NSFC) No. 11273005, SRFDP (20120001110064), and the 973 Program (2014CB845800).

\end{acknowledgements}

\end{document}